\documentclass[%
 pre,twocolumn,
 amsmath,amssymb,
 aps,
]{revtex4-1}

\pdfoutput=1

\usepackage{graphicx}
\usepackage{dcolumn}
\usepackage{bm}

\begin{document}


\title{Extracting Hidden Hierarchies in 3D Distribution Networks}

\author{Carl D. Modes$^1$, Marcelo O. Magnasco$^1$, and Eleni Katifori$^2$}
 \affiliation{$^1$The Rockefeller University, New York, NY 10065}
 \affiliation{$^2$Max Planck Institute for Dynamics and Self-Organization (MPIDS), 37077 Goettingen, Germany.}
 \email{cmodes@rockefeller.edu}

\date{\today}

\begin{abstract}
Natural and man-made transport webs are frequently dominated by dense sets of nested cycles. The architecture of these networks, as defined by the topology and edge weights, determines how efficiently the networks perform their function. Yet, the set of tools that can characterize such a weighted cycle-rich architecture in a physically relevant, mathematically compact way is sparse. In order to fill this void, we have developed a new algorithm that rests on an abstraction of the physical `tiling' in the case of a two dimensional network to an effective tiling of an abstract surface in space that the network may be thought to sit in. Generically these abstract surfaces are richer than the flat plane and as a result there are now two families of fundamental units that may aggregate upon cutting weakest links -- the plaquettes of the tiling and the longer `topological' cycles associated with the abstract surface itself. Upon sequential removal of the weakest links, as determined by the edge weight, neighboring plaquettes merge and a tree characterizing this merging process results. The properties of this characteristic tree can provide the physical and topological data required to describe the architecture of the network and to build physical models. The new algorithm can be used for automated phenotypic characterization of any weighted network whose structure is dominated by cycles, such as mammalian vasculature in the organs, the root networks of clonal colonies like quaking aspen, or the force networks in jammed granular matter. 

\end{abstract}

\pacs{Valid PACS appear here}
\maketitle


\newcommand{\e}{\cdot 10^} 
\newcommand{\dd}{\mbox{d}}

\section{Introduction}

Complex networks are a pervasive presence in both modern technology and the natural world, with examples as widely varied as transportation networks, the internet, mammalian vasculature, the neuronal connections in the brain, load-bearing architecture, river deltas, venation in plant leaves, and even slime molds. Driven largely by the explosion in importance of the world wide web, academic interest in the associated study of complex networks has risen in prominence as well in recent years, with the introduction of important new classifying ideas such as "small-world" \cite{Watts1998} and "rich-get-richer" \cite{Barabasi1999} network topologies. Inspired by the anything-goes hyperlinked structure of the internet itself, many of these studies concentrate only on the topological and graph-theoretic properties of general classes of complex networks with unrestricted node degree.

On the opposite end of the spectrum from these hyperlinked networks are those systems who naturally live in two dimensions, such as river deltas or plant leaf venation. Much recent work has also served to elucidate matters here as well, from a comprehensive sedimentary modeling of the river delta structure and evolution \cite{Seybold2007, Iturbe1997} to the construction of a hierarchy-sensitive and geometry independent topological characterization of the plant leaves \cite{Katifori2012, Mileyko2012}. Other methods that rely on local geometric cues to characterize leaves, road networks and crack patterns \cite{Perna2011}, or topological characterization of networks of epithelial contacts \cite{Escudero2011} have also been explored.

There remains, however, an important class of complex networks that lie somewhere in between these two extremes, sensitive to geometric embedding in space yet too complex in structure to live entirely in the plane. This intermediate class has seen relatively fewer recent advancements, with most activity centered around mapping and descriptive efforts of certain functionally relevant network phenotypes \cite{Schaffer2006, Blinder2010} though even simple vascular cartography is fraught with difficulties at the mesoscale, with a complete map of a mouse liver only emerging recently \cite{Oppenheim2014}. Several members of this intermediate class are networks where a deeper understanding would have far reaching biomedical implications, such as vascular or neural nets, so the impact of a proper tool to characterize them would be immense, opening the door to predictive modeling of function and disease-process-driven malfunction \cite{Fernandez2009}. Furthermore, powerful and effective quantitative descriptors are a necessary first step to both disentangling the network architecture's role during development -- of itself and the surrounding tissue -- and to advanced, three-dimensional artificial organ synthesis.

The idea behind our cycle-hierarchy sensitive characterization is based on one of the recent methods developed to describe distribution networks in two dimensions \cite{Katifori2012} -- where adjacent cycles in the network are allowed to sequentially merge into larger and larger cycles. The order of merger and relative location of these cycles can be mapped onto a bifurcating tree graph and a number of simple statistical measures of this tree are then easily accessible and contain information about the original network. For a distribution network -- or, more generally, any weighted network -- that lives in three dimensions, however, we are presented with an imposing hurdle at the very beginning of the process: what does it mean for two cycles to be adjacent? Our method solves this question by imagining the network lives on a topologically non-trivial surface in space and represents an effective tiling of that surface, from which cycle adjacency and the rest of the prescription from the two-dimensional case naturally follows.  

We begin in Section II with a quick primer and refresher on some classical results of graph theory necessary to understand the core of our algorithm. We then lay out the machinery of our characterization and quantification algorithm in Section III, and we use it on several examples of abstractly interesting or physically relevant surrogate networks in Section IV. We discuss implications and future directions in the concluding Section V.  

\section{A Brief Graph Theory Refresher}

In the interest of providing a quick reference and explanation of some of the classical graph theoretic concepts integral to this work, this section will  cover the concepts of a graph's cycle space and graph embeddings and genus. This section is intended as a simple review; a reader comfortable with these topics should feel free to skip ahead. Our new cycle-coalescence algorithm is presented in Section III.

\subsection{The Cycle Space}

Since our ultimate target is to construct a characterization of cyclic paths in distribution networks it will be helpful to make contact with the mathematical structures that cycles in a graph may be endowed with. For the remainder of this work, we assume that all graphs are simple graphs, i.e. no pair of vertices may be directly connected by more than one edge. Given an arbitrary such graph, it turns out that the superset of all cyclic paths without repeated traversal of the same edge can be thought of as a vector space, known as the \textit{cycle space} \cite{Veblen1912}. In this formalism, the individual elements (i.e. the `vectors') are sets of cyclic paths represented as the entire graph with each edge assigned a $0$ if it is not traversed in a cycle and a $1$ if it is. The dimensionality of this vector space is accordingly less than the number of edges in the graph. The vector addition is defined as a simple edgewise addition of the graph representatives for each vector being added. Furthermore, the cycle space in its simplest form is defined over $\mathbf{Z}_2$ so if an edge is used in each member of the binary summand it is \textit{not} used in the sum. Thus, for example, if one were to add two adjacent cycles in a graph, the result would be the larger, boundary cycle. 

Why do elements defined on a graph in this way, equipped with this addition rule constitute a vector space? Much of this is down to the simplicity of working over $\mathbf{Z}_2$ -- scalar identity, compatibility between the multiplication and addition in $\mathbf{Z}_2$, and distributivity of the scalar multiplication with respect to both the vector addition and the field addition all follow trivially from $\mathbf{Z}_2$'s simple two-element structure. Meanwhile, the associativity and commutativity of vector addition are satisfied by construction, leaving only the specification of an identity element and inverse elements under the vector addition for any member of the vector space. Clearly a graph adorned with zeroes on every edge, a null cycle, is the appropriate zero vector. Again appealing to the simplicity of $\mathbf{Z}_2$ it is clear that every vector in the cycle space is its own inverse since each edge with a zero will remain zero and each edge with a one will be set to zero upon the addition of a vector to itself. Note that, further owing to that same simplicity of working over $\mathbf{Z}_2$, the cycle space is \textit{not} equipped with an inner product.

Having established that the cycle space is truly a vector space it is a natural next step to discuss some of its relevant properties. As all vector spaces must be, the cycle space is equipped with a basis and as it happens, there also exists a convenient way to generate such a basis. By choosing any spanning tree -- a subgraph that contains every vertex but no cycles \cite{Gross2005} -- on the graph and adding a single further edge to the tree one sees that precisely one cycle is created by the union of the spanning tree with the chosen extra edge after pruning all extraneous tips (or `leaves') left over from the spanning tree. Furthermore, choosing a different extra edge produces a different cycle. The set of all cycles created in this way constitutes a basis for the space \cite{Kocay2004}. Clearly, if one had chosen a different spanning tree to begin this process then a different basis would result, but despite the large number of spanning trees available for an arbitrary graph -- typically exponential in the number of vertices and possibly worse \cite{Kocay2004}  -- bases generated in this way are referred to as \textit{fundamental cycle bases} \cite{Veblen1912} and are actually quite special and in some cases comprise only a small portion of the total number of possible bases. Note that this procedure provides a direct way to compute the dimension of the cycle space: for a graph with $v$ vertices and $e$ edges, a spanning tree must use precisely $v-1$ edges, leaving $e-v+1$ edges to create the fundamental cycles. Hence the dimension of the cycle space for an arbitrary graph is $e-v+1$.

Furthermore, among this massive collection of possible bases for the space, one may look for special bases that optimize certain simple properties. For example, there exists a \textit{minimum weight basis} which contains the collection of basis vectors whose combined sums over the edge weights is as small as possible. In the event that every edge is simply assigned a weight of $1$ then the minimum weight basis corresponds to the basis which collectively uses as few edges as possible. This particular minimum weight basis suggests a connection to tilings -- if a graph is a simple tiling of the plane, such as a checkerboard, then it is easy to see that not only is the set of individual tiles a basis for this graph's cycle space but that this basis is minimum weight in the way described above. There is one subtlety here that must be pointed out, however -- if the boundary cycle (i.e. the cycle that remains upon summing all the tile basis vectors) has fewer edges than one of the actual tiles, then it is included in the minimum weight basis in the place of the largest tile. There is a well-defined sense in which it is actually more appropriate to think of these plane tilings as tilings of the sphere instead, wherein the boundary cycle is just another tile. In this setting, the minimum weight basis is exactly the set of all the tiles but the largest. To push this analogy further, though, we must first confront the separate subject of graph embeddings. 

\subsection{Graph Embeddings}

\begin{figure}[t]
\includegraphics[width=0.95\linewidth]{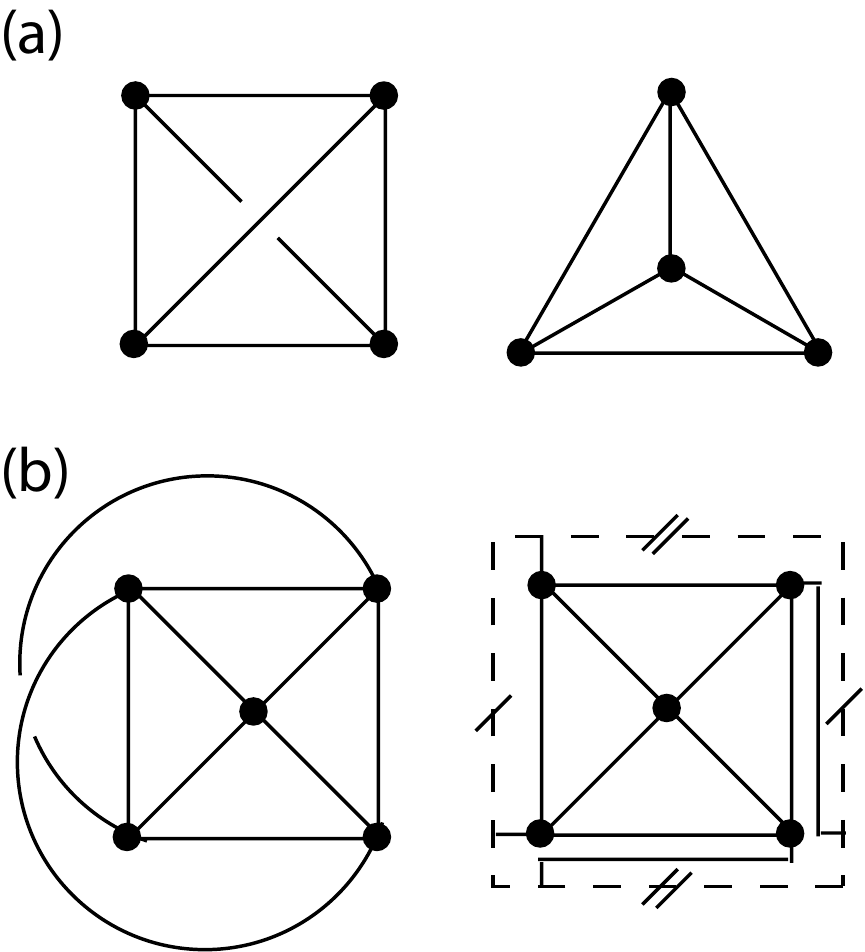}
\caption{\label{fig:graphEmbed} Some simple examples of graph embeddibility, both planar and otherwise. (a) The complete graph on four vertices, $K_4$, is not embedded in the plane in its most traditional representation (left), but is planar nonetheless (right). (b) The complete graph on five vertices, $K_5$ is not planar (left). However, this graph can be represented without edge crossings on a torus (right).}
\end{figure}

As outlined above, certain special graphs may be thought of as tilings of the plane. This occurs when a graph may be represented in the plane without any edges crossing one another; graphs that have this property are known as \textit{planar} graphs. All representations of a planar graph do not necessarily sit in the plane without edge crossings -- just one such representation will suffice. As an example, the complete graph on four vertices, $K_4$, is not embedded in the plane in its most traditional representation, but is planar nonetheless (see Fig. \ref{fig:graphEmbed}a). A randomly chosen graph with many vertices, however, is vanishingly unlikely to be planar \cite{Gimenez2009} -- indeed one only needs five vertices for non-planar graphs to begin to appear as $K_5$ is one such, and they rapidly dominate as more vertices are added.

On the other hand, all graphs may embed into space with no edge crossings. The simplest way to see that this is true is through a process known as the \textit{book embedding}\cite{Bernhart1979}: imagine placing all vertices on a line in space, to be thought of as the spine of an open book. Then every edge may be placed connecting its two end-point vertices without the possibility of intersecting any other edge if each edge has a page of the book to itself. Since such an embedding is always possible it follows that embedding a graph without edge crossings on a complicated surface is also always possible -- simply construct a surface with characteristic distances smaller than the vertex separations and follow the edges of the book embedding. This surface is, of course, extremely complicated, with genus, $g$ -- the number of ``holes" or ``handles" in the surface -- of order the number of edges in the graph. However, since a graph may always sit in a surface in space it is well-defined to ask what is the simplest (i.e. lowest genus) surface that can accommodate an embedding of a particular graph. As an example, consider again the simplest graph that is not planar, $K_5$ -- as shown in Fig. \ref{fig:graphEmbed}b, this graph can be represented without edge crossings on a torus. Unfortunately, in general the problem of determining the graph genus for an arbitrary graph is NP-hard \cite{Thomassen1989}. However, the existence of a simplest topological surface on which a graph may be thought to tile is all that is needed to proceed.

\begin{figure}[t]
\includegraphics[width=0.95\linewidth]{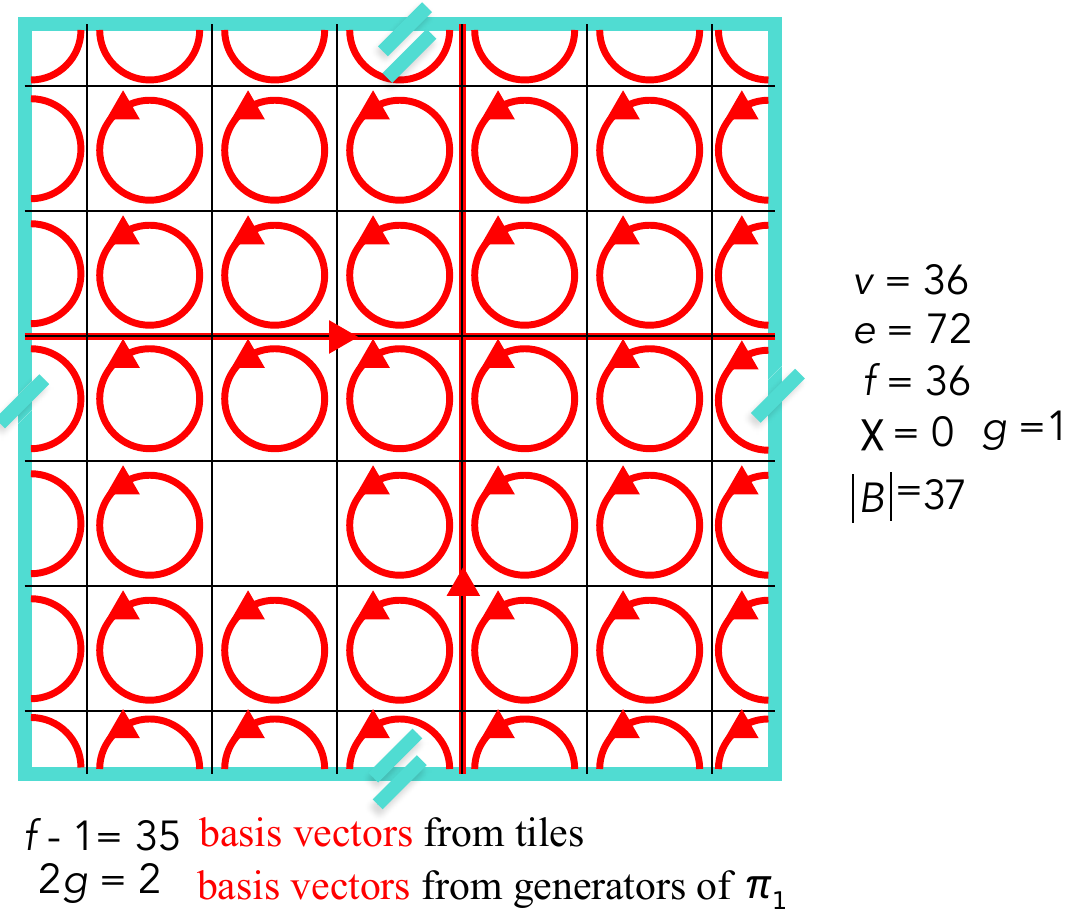}
\caption{\label{fig:cyclespace} Tiles and topological cycles of a square lattice on a toroidal topology (g=1). Opposite sides of the lattice are identified as shown.}
\end{figure}

Since planar tilings represented a simple example in the previous discussion of the cycle space, it is worth pointing out what happens in the more general, non-planar case. As just discussed, \textit{every} graph may be represented as a tiling on a surface of some genus and clearly a minimum weight basis on the edge count must at least include all the tiles (but the largest one, for the same reasons as above). But is there anything new? We know that the number of elements of the basis must be $e-v+1$, and we also know that the number of elements that came from tiles is $f-1$ for a tiling with $f$ faces or tiles. Putting these two facts together with the formula for the Euler character in terms of the genus, $v-e+f=2-2g$, yields a clear overage due to the topology:
\begin{equation}
e-v+1 = (2g)+(f-1)
\end{equation}
Indeed, the promotion of the embedding surface to one with genus necessitates some new basis elements: the generators of the fundamental group of the surface \cite{Poincare1895, Hatcher2002}, which encodes the contractibility of families of paths on the surface. There are precisely $2g$ such generators (see Fig. \ref{fig:cyclespace}). As can be seen by inspection in Fig. \ref{fig:graphEmbed}b or Fig. \ref{fig:cyclespace}, no sum of the simple tiles can give a (single) cycle that has nonzero winding across one of the two periodic boundaries, and yet these cycles clearly exist in the graph so the cycle basis must generate them somehow. The inclusion of these new elements resolves this issue.

We therefore expect our cycle basis to consist of $f-1$ tiles and $2g$ topology generators. For graphs that can be embedded on toroidal surfaces with `fat' handles, the sum of edges of the $2g$ topology generators should be all greater than the sum of edges of the tiles, thus allowing us to easily identify the tiles from the generators. However, such a clear size separation of the tiles and generators is not always guaranteed: otherwise, direct reconstruction of the graph embedding would be possible in polynomial time. We construct our algorithm and discuss some of the expectations and statistics related to this non-guarantee and the corresponding ability to extract information without explicitly constructing the embedding in the next section.

\section{The Cycle-Coalescence Algorithm}

We now describe how the vectors of the minimal basis are hierarchically added by the cycle coalescence algorithm. The algorithm consists of three separate parts, and eventually generates a characteristic, linkage tree. We will assume that the networks that are being analyzed have no loops (edges that terminate at the same node), double edges or bridges, i.e. that they constitute simple graphs that are at least 2-connected.

First, we describe how to obtain the minimum weight basis over the edge count of the unweighted graph, hereon termed {\it minimal basis} to avoid confusion. Second, we show how to sequentially merge the identified tiles and construct the characteristic tree. Last, we discuss various metrics one can use to describe the structure of the characteristic tree, and what these metrics mean for the architecture of the original graph.

\subsection{Finding the Minimum Weight Cycle Basis}
\begin{figure}[t]
\includegraphics[width=0.8\linewidth]{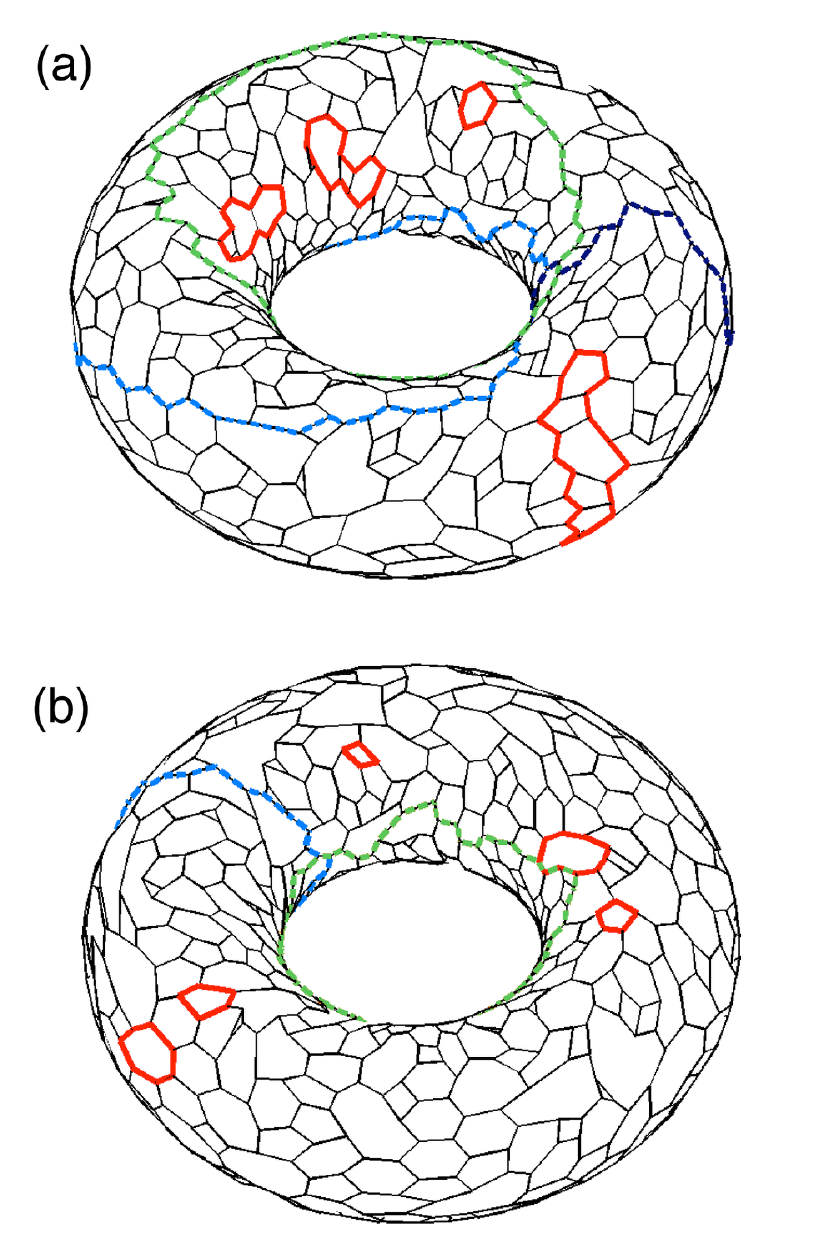}
\caption{\label{fig:torus1} A graph embedded on a toroidal surface. Highlighted are example basis vectors from (a) a fundamental cycle basis and (b) a minimum weight basis. Red: some representative tiling basis vectors. Blue and orange: basis vectors that correspond to the $2g=2$ generators of the fundamental group of the torus. }
\end{figure}

\begin{figure}[t]
\includegraphics[width=0.95\linewidth]{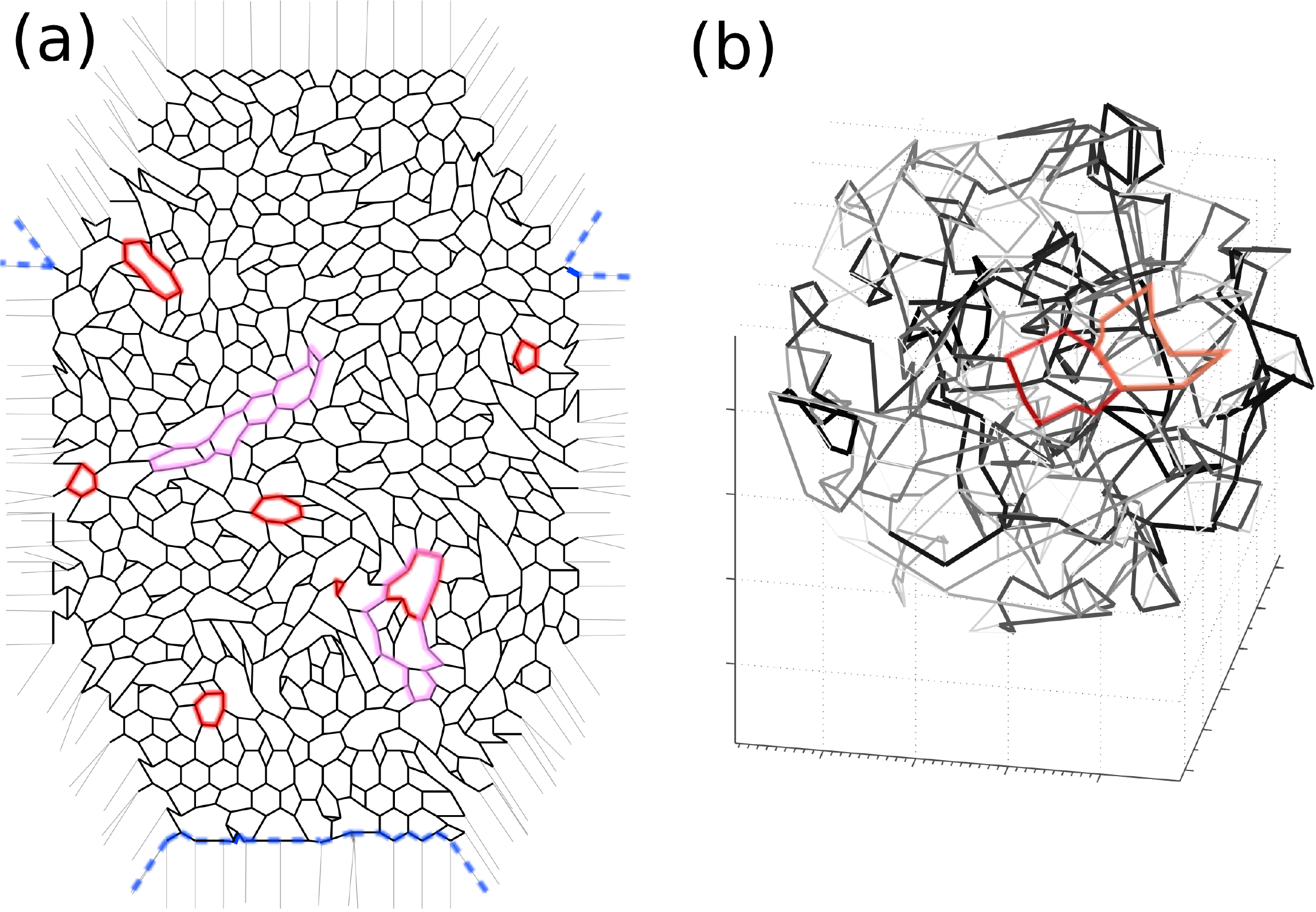}
\caption{\label{fig:torus2} (a) Graph embedded on a two-holed torus, schematic. Periodic boundary conditions connect opposite sides of the octagon. Highlighted are some example tiling basis vectors from a fundamental cycle basis (magenta) and from the minimum weight basis (red). The blue curve highlights one of the $2g=4$ vectors in the minimal basis corresponding to the generators of the fundamental group of the two-holed torus. (b) A 3D graph with an unknown embedding. Identification of tiling versus generator basis vectors is not straightforward any more. Highlighted with red and orange are two vectors in the minimum weight basis that share an edge.}
\end{figure}

There are a number of polynomial time algorithms that can be used to construct the minimum weight basis \cite{Horton1987}. Starting from an arbitrarily chosen node in the unweighted network, we find the minimal spanning tree -- equivalent to executing a breadth-first search over the graph -- and determine the fundamental cycle basis associated with that node. We repeat for every node of the network, in this way generating a set $S$ containing all the unique vectors of the fundamental cycle bases associated with each node. We sort the set $S$ based on vector size in an ascending order and start building a minimal basis $B$ bottom up, by sequentially adding vectors from smallest to larger. When two vectors in $S$ are degenerate, their respective order in the sorted set is determined at random. Before any vector $b_i$ is added to $B$, we check if that vector is linearly independent from the vectors already added to the set. If the vector depends linearly on the vectors already added to the set, that vector is discarded and the next vector $b_{i+1}$ is checked for linear independence. The set $B$ closes and becomes a real basis when no new vector can be added. However, we do not need to check every vector in $S$ - we know that the number of elements in the basis is $e-v+1$, so the search is terminated when the cardinality of $B$ becomes $e-v+1$. Examples of fundamental basis vectors in $S$ and minimal basis vectors in $B$ can be seen in Fig.~\ref{fig:torus1} and in Fig.~\ref{fig:torus2}(a). 
In the case of degeneracies, i.e. the existence of more than one minimum basis, the output minimum basis might be dependent on the chosen algorithm. However, the statistical properties of the final characteristic tree are generally robust and, with the exception of some singular cases, do not depend sensitively on the exact basis. Furthermore, we expect that most of the degeneracy will be confined to cycles representing topology generators not tiles, further minimizing their impact on the final output.

\subsection{Constructing the Characteristic Tree}
The minimum weight basis $B$ contains only topological information of the original graph, as in constructing $B$ we ignored the weights of the individual edges. The cycle coalescence algorithm integrates the structural information carried by the edge weights in the construction of the characteristic tree. The algorithm begins with a set of vectors $B_{iter}$ identical to the basis $B$. The set $B_{iter}$ will be updated (vectors added and removed) throughout the algorithm

\begin{figure}[t]
\includegraphics[width=0.95\linewidth]{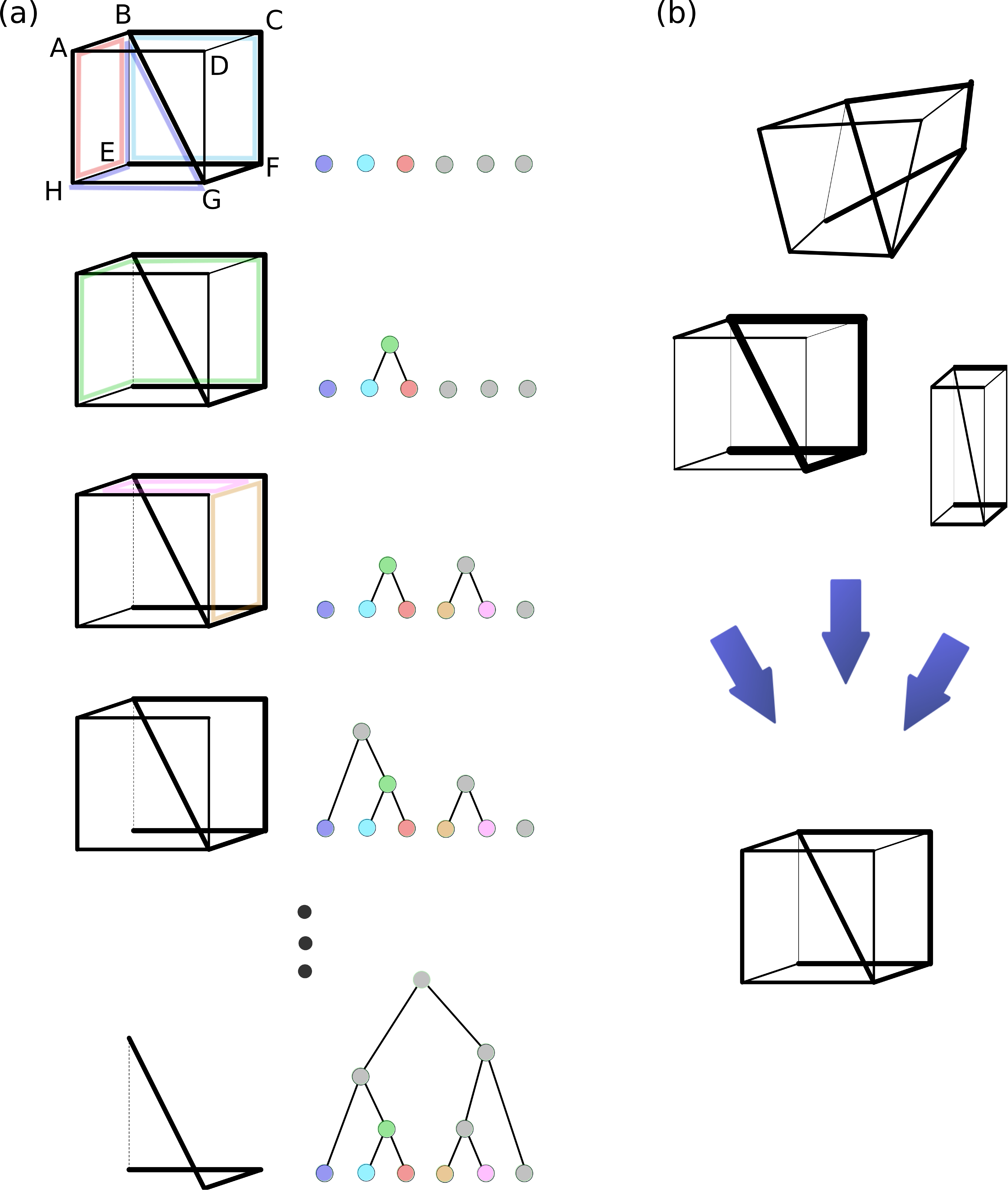}
\caption{\label{fig:method} (a) Schematic of the cycle-coalescence algorithm applied to a simple 3D graph. The cycle coalescence proceeds from top to bottom. On the right we show the characteristic tree being progressively built. To aid the eye, some cycles that participate at that stage of the algorithm are highlighted in color, and the corresponding nodes of the characteristic tree are correspondingly color coded.  (b) Schematic of degenerate graphs that produce the same characteristic tree. The cycle coalescence is blind to geometry and exact weights of links.}
\end{figure}

To begin, we identify the edge $e_i$ of the graph with the smallest weight. We then locate the basis vectors $\{b_k\}$ in $B_{iter}$ that contain that edge. An example is shown in Fig.~\ref{fig:method}(a), where basis vectors \textit{BEFC}, \textit{BEHA} and \textit{BEHG} all pass through edge \textit{BE}. The two shortest vectors that contain that edge are added as described in Section II, creating a new cycle. As described below, this choice will statistically preferentially merge tile cycles over generator cycles, creating a steadily-ever-more coarse-grained tiling of the abstract surface. This new cycle is added to $B_{iter}$, and the two original vectors are removed. In our implementation degeneracies are again resolved here by random choice within the degenerate vectors. In Fig.~\ref{fig:method}(a) the addition of \textit{BEFC} and \textit{BEHA} results in cycle \textit{FCBAHEF}, color-coded in green. The algorithm proceeds by identifying the next smallest edge that is utilized by at least two vectors in $B_{iter}$ and iteratively repeating the process until there is only one cycle left, as in the bottom of Fig.~\ref{fig:method}(a).

This cycle-coalescence algorithm can be represented with an unweighted, bifurcating tree whose nodes represent cycles of the original graph and the links connect cycles that are connected via cycle addition operations. In particular, the basis vectors in the initial set $B$ are the leaf (terminal) nodes of the tree, represented by the six nodes on the top right panel of Fig.~\ref{fig:method}(a). When two cycles are added, they result in a new cycle, represented by their {\it parent} node in the bifurcating tree. Hierarchically joining the cycles based on the sort order of the edges thus results in a tree whose bifurcation statistics capture the structure of the nested cycles of the network. An architecture dominated by highly nested cycles will produce a close to perfect binary tree, whereas a more disordered architecture will produce an asymmetrical tree \cite{Katifori2010}. 

The information encoded in this characteristic tree does not depend on the geometric location of the nodes, nor the exact value of the edge weights, but only depends on the network connectivity and sort order of the edge weights. In fact, as shown in the example of Fig.~\ref{fig:method}(b), the characteristic tree is invariant under any node movement or edge weight change (provided that the sort order is maintained). The characteristic tree is thus an ideal tool to describe structural information about the network not captured by widely used metrics such as edge weight distributions or weighted or unweighted degree distributions \cite{Newman2003, Barrat2004}.

The computational complexity of the algorithm is dominated by finding the minimum weight basis. An optimized implementation is capable of running in polynomial time $O(e^2 v / \log v)$.

\subsection{The Statistical Arguments for the Cycle-Merger Choice}

In the above described cycle-coalescence algorithm we have assumed that the appropriate choice of cycle space vectors to merge upon the cutting of an edge in the graph is simply the two shortest cycles by number of edges traversed. The rationale for making this choice is that in the abstract representation of the graph as a tiling of a potentially complex surface living in three dimensions we desire to merge adjacent tiles rather than cycles that have arisen due to the non-trivial topology of the surface. The intuition is that, generically speaking, `real' tiles should be shorter than cycles that traverse entire `handles' of the surface, especially as the number of vertices in the graph tends to infinity for controlled degree distributions. This intuition can be made concrete in a statistical sense, but to do so we must consider how a basis vector of the cycle space that arises from topology and not from tiling might be shorter than one that represents a tile and how likely is it that this occurs. In order to ensure that the graphs we consider comply with the need for a controlled degree distribution, we restrict ourselves in what follows to the simplest such family: $3$-regular graphs. Despite this choice, we do not lose much for relevance as nearly all biological and most physical distribution networks are of this type, having developed through multiple stages of binary branching and/or tip-to-channel reconnections \cite{Risau1997}. Additionally, the form if not the specific detail of the following arguments will apply to other families of graphs with different, controlled degree distributions.

Before proceeding, it is useful to establish a `basal' $3$-regular tiling from which we can explore more complicated variants. We will leave the genus of the surface, $g$, as a free parameter. A tiling of such a surface satisfies the familiar formula for the Euler Character in terms of the genus invoked earlier: $2-2g = v - e - f$. Furthermore, for a $3$-regular graph, $v$ and $e$ are related by $e = 3v/2$. How does the genus affect the average number of edges per tile, $|p|$? Since every face shares each of its edges with another face, we must have $|p|f=2e$ and hence:

\begin{equation}\label{eq:sides}
|p| = \frac{6v}{4+v-4g}.
\end{equation}

\begin{figure}[t]
\includegraphics[width=0.95\linewidth]{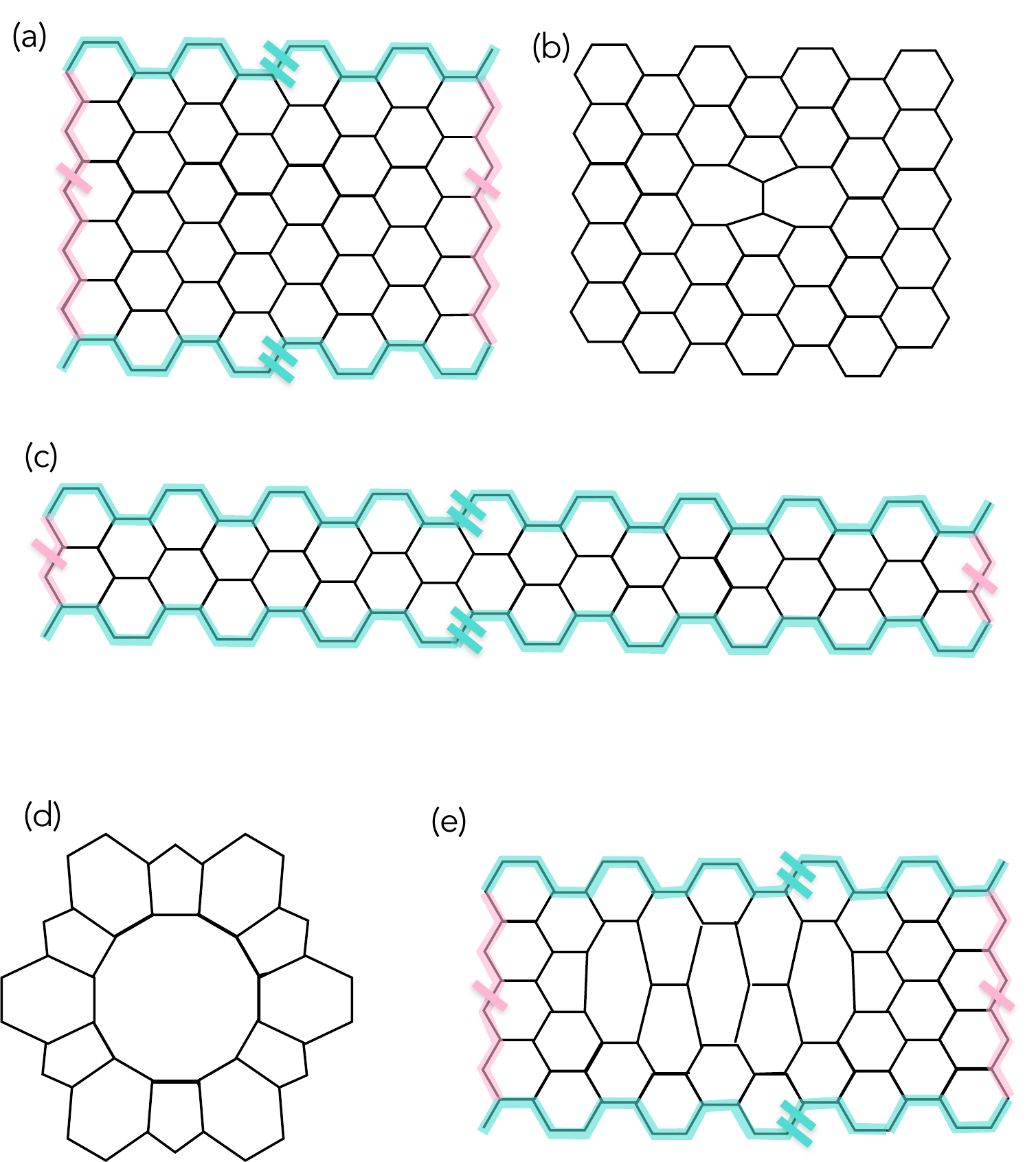}
\caption{\label{fig:statarg} (a) Honeycomb lattice on a toroidal topology. Opposite sides are identified as shown. (b) T1 topological rearrangement on a honeycomb lattice. (c) One holed-torus with a thin handle. (d) A tile with a large number of sides as a result of condensation of disclination charges. (e) Varicosity as a result of a condensation of like-charged dislocation defects. Opposite sides are identified as shown, resulting in the topology of a one-handle torus.}
\end{figure}

It is therefore clear that so long as the number of vertices in the graph is much larger than the genus of the tiled surface we may imagine our basal tiling as a simple hexagonal net with a handful of isolated heptagonal (necessary by Eq.~\ref{eq:sides} when $g>1$) or pentagonal tiles (when $g=0$) due to the topology of the surface. As an aside, in the case of a spherical topology this is the source of the icosahedrally symmetric pentagonal defects one encounters in, for example, viral capsids, fullerene, geodesic domes, and soccer balls \cite{Coxeter1962}. Note that for a simple toroidal topology with $g=1$ the basal tiling is a perfect hexagonal crystal (Fig. \ref{fig:statarg}a).

Having established our set of basal tilings, we must consider the fundamental unit of perturbation away from these tilings that still respects the topological relationship discussed above. In particular, we may not freely insert tiles with more or fewer sides than the average given the genus and number of vertices -- we must at the least replace a pair of hexagons from the basal tiling with a pair of polygons whose total number of sides sums to $12$ to ensure we satisfy the necessary $|p|$. A positive departure from 6-sidedness in this context is known as a negative \textit{disclination defect} and a negative departure a positive disclination defect. In this language, the necessary condition of a perturbation is that it has no net disclination charge. This condition is not, however, sufficient: a disclination dipole produces a dislocation defect on this crystalline background and since our tiling exists on a compact, closed surface without boundary it cannot escape to infinity nor can it simply end on a boundary -- it must be terminated by a second disclination dipole. These conditions together, that there is neither disclination nor dislocation charge are necessary and sufficient for the production of a self-consistent perturbation. These perturbations take the form of a disclination quadrupole also known as a T1 topological rearrangement \cite{Weaire2000} (Fig. \ref{fig:statarg}b).

With our basal tilings and working perturbative units in hand, we may now tackle the question before us: how can the assumption that tiles are ``shorter" than topological cycles fail? One possibility is that one of the directions around a handle is very short (Fig. \ref{fig:statarg}c). For fixed $v$, a surface may be represented as a $4g$-gon with appropriate sides identified \cite{Sausset2007}. In the case in which all handles are of the same thickness and length, this $4g$-gon becomes regular, with side lengths that scale as $v^{1/2}/g$. Unless $g$ is very large relative to and scales linearly with $v$, the number of cases where a shrinking handle is still `thicker' than the circumference of a tile is of order $v^{1/2}$ and hence a problem of this type is encountered with likelihood that scales as $1/v^{1/2}$. Note that an extrinsically growing genus does represent a real issue here, and we conclude that our algorithm is less likely to provide meaningful insight for ordered structures with small, simple unit cells since the genus in these cases is large relative to $v$ and grows linearly with the addition of further unit cells to the crystal.

Turning to our quadrupolar perturbations, we see that another potential problem arises not when a handle traversal becomes short, but when a tile becomes large: condensation of one of the disclination charges upon repeated perturbations can lead to a tile with a large number of sides (Fig. \ref{fig:statarg}d). With $f=2e/|p|$ faces in a $3$-regular graph let us further assume that the genus is low enough relative to $v$ that $|p|$ is near $6$, that is, that the basal tiling is still a hexagonal net with sporadic heptagons. Since we have already discounted very high genus scenarios due to the above reasoning for handle thin-ness cases, little further is lost here. With $|p| \approx 6$, we have $f \approx v/2$ and $e = 3v/2$. There are precisely as many T1 topological rearrangements available as there are edges in the graph and those that increase the number of sides of a given tile occur by selecting an edge that shares a vertex with the tile, but is not an edge for it. Since the graph is three regular, there is only one such edge per vertex. Therefore the number of perturbations that increase the number of sides for a given tile with $p$ sides is simply $p$. Note that the same number of perturbations exist that decrease the number of sides for that tile. The likelihood of significant condensation is clearly very low, occurring at a rate:
\begin{equation}
P(p=N>6) = \frac{2 N!}{7! \left( \frac{3v}{2} \right)^{N-7}} 
\end{equation}
for $N-6$ perturbations and may be safely ignored.

Finally, there is the possibility that a handle experiences varicosity. Unlike in the first case considered where the handle thickness is simply too small across the entire handle, when a handle experiences varicosity the handle thickness varies as a result of a condensation of like-charged dislocation defects and becomes locally too thin as a result (Fig. \ref{fig:statarg}e). $L$ consecutive T1 events must occur in exactly the right position -- overlap of neighboring $5-7$ pairs -- simply to open up a single dislocation `scar' of length $L$, with likelihood scaling as $1/v^L$. This must occur several more times \textit{and} the opened scars must align \textit{and} their projection onto the direction normal to the handle traversal must result in concurrent overlap for there to be any chance of generating a cycle shorter than the hexagonal and heptagonal tiles that are the primary constituents of the graph. This eventuality, too, may be safely ignored.

Since, as we discuss in the following subsection, the characteristic tree will ultimately be subject to its own round of statistical analysis there is even more built-in statistical robustness than even the above arguments indicate. Even if the characteristic tree fails to accurately recapitulate the `true' tile coalescence pattern, this failure will be isolated to a handful of specific nodes and will have minimal effect on statistical measures of the tree.

\subsection{Quantifying the Characteristic Tree}

The characteristic tree provides a convenient way to analyze the architecture of a complex weighted network that is composed of cycles.
The topology of the characteristic tree reflects the hierarchy of cycle nesting in the original weighted graph. A detailed discussion on how different tree bifurcation statistics map to various graph architectures and how one can use the characteristic tree to quantify the degree of nestedness of planar graphs and to analyze their weighted topology, can be found in \cite{Katifori2012}. In summary, the more balanced (low height) the tree is, the more highly nested the original graph. High weight cycles are subdivided iteratively by smaller weight edge, creating a cascade of hierarchically nested cycles. On the other hand, high height, unbalanced trees typically represent graphs where smaller weight cycles are added sequentially on the backbone of bigger cycles, creating an architecture that is less hierarchically organized.  

To analyze the characteristic tree, we need to assign a number to each tree architecture. There are several schemes that have proven quite useful in quantifying several aspects of the tree: Horton and Strahler numbers, the tree height etc \cite{Horton1945, Strahler1952, VanPelt1992}. Each scheme has its advantages and disadvantages, and discriminatory power that focuses on different aspects of the architecture.  In this paper, we use an adapted version of the partition asymmetry, as introduced in \cite{VanPelt1992}. The partition asymmetry is a metric that characterizes the overall topological structure of a binary tree, and quantifies the difference in size (number of leaf nodes) between the two subtrees that stem from a tree vertex. We define the partition asymmetry $a(j)$ of a bifurcation vertex $j$ as: 
\begin{equation}
a(j)=\frac{s_j-r_j}{s_j+r_j-1}
\end{equation}
with $s_j>r_j$ and $s_j+r_j\ge 2$. The parameters $r_j$ and $s_j$ are the degrees of the two subtrees at partition $j$. The degree of a (sub)tree is defined here as the total number of the leaf nodes (terminal segments) of that (sub)tree. 

The partition asymmetry $a(j)$ provides a number for each vertex of the characteristic tree that quantifies the degree of hierarchical organization in the loop nestedness of the original graph. A tree that is balanced and corresponds to a hierarchically nested graph will have many low asymmetry vertices. Conversely, a graph that is not hierarchically nested will produce a characteristic tree with many vertices that have high asymmetry. Thus, comparing the distribution of partition asymmetries of two graphs can be a metric on how statistically similar the architecture of those two graphs is. 

In order to characterize the architecture of the graph, instead of pairwise comparison of asymmetry distributions, we define the topological asymmetry $A$ of a graph, by measuring the percentage of vertices in the tree that have asymmetry higher than $0.95$. 
\begin{equation}
A=p(j | a(j)>0.95)
\end{equation}
 The higher the topological asymmetry of the graph, the less hierarchically nested the graph is.

Each node in the characteristic tree represents a cycle in the original graph, and the subtree that is rooted at that node encompasses information about the architecture of part of the graph ``contained" in the cycle. The higher the degree of the node in the characteristic tree, the bigger the part of the graph represented in the subtree. Thus, as the tree contains many more nodes far away from the root than close to it, the distribution of the partition asymmetry is dominated by the architecture at the small length scales. There are many more nodes close to the leaves of the characteristic tree, meaning that any partition asymmetry average, unless weighted in favor of high degree nodes, will be dominated by the architecture at the short length scale. If the architecture of two graphs is statistically similar at small length scales, the topological asymmetry $A$ will not have enough discriminatory power to distinguish them. We need a second quantity that characterizes some of this missing information.

The average vein length $ L $ is a topological length that quantifies the average length of the ``veins" in the weighted graph (\textit{not} the characteristic tree itself). It is calculated by constructing trails where the edge weight declines monotonically. Starting from an initial link $e_1\equiv \langle i_1 j_1 \rangle$ between nodes $i_1$ and $j_1$ (first link of the trail $t=( e_1)$), we identify all the links $\{e\}$ that are adjacent to it (share node the $i_1$ or $j_1$) and have weight smaller or equal to the weight $w(e_1)$ of edge $e_1$ . We choose the link with the maximum weight from the set $\{e\}$, which we will call $e_2$, and add it to the trail, which now becomes $t=( e_1, e_2)$. We now repeat the process for $e_2$ (identify all links that are adjacent to $e_2$ with weight smaller than $w(e_2)$ and chose the maximum) and iterate. The algorithm is stopped when the set of the links that have weight smaller than the weight $w(e_k)$ of the last link $e_k$ in the trail is empty. The length of the trail $t=(e_1,e_2,...,e_k)$ associated with edge $e_1$ is $l(e_1)=k$. We repeat the process starting from every link of the graph, and this way, associating a trail length  $l(e)$ with every link $e$. 
The average vein length is defined as 
\begin{equation}
 L  = \frac{1}{|e|}\sum_e l(e).
\end{equation}
Note that it is generally not the case that $ L $ and $A$ may be varied independently by picking edge weights ``by hand," especially if the embedding structure is unknown.

Here we need to stress that the topological length  $ L $ and the topological asymmetry $A$ are just two of the many ways one can measure the architecture using purely topological information. Equivalent choices could be weighted asymmetries with weights that favor nodes close to the root of the characteristic tree, for example -- as pointed out above, the statistical tools available for analysis of a tree graph are manifold. The best choice can and should depend on the nature of the data being analyzed.

\section{Examples and Surrogate Data}

To test our new characterization tool we computer generated a series of weighted networks produced by a number of distinct generation protocols. We chose these generation protocols so that the produced networks would be statistically indistinguishable under many widely used network metrics, to demonstrate the power of the cycle-coalescence algorithm. We also chose generating functions with some biological relevance to vascular networks. The degree of all the networks was strictly equal to three, as would typically be the case in a natural transport network -- as pointed out above -- such as plant or animal vasculature \cite{Risau1997}. 

\subsection{Example Networks}

For our example networks, we considered two types of underlying topologies. In the first, the graph by construction could naturally be embedded on a two holed torus. The original graph was produced by generating a regular hexagonal grid on a plane, applying the appropriate periodic boundary conditions to create a two-holed torus and finally applying a random series of T1 transformations to introduce lattice defects and randomize the graph. An example is shown in Fig.~\ref{fig:torus2}(a).

In the second, the graph was generated by progressively joining nearby nodes randomly scattered inside a 3D sphere, so that each node has a maximum of 3 links. 
To begin, we randomly distributed $N$ nodes inside a 3D sphere. We identify the node closest to the center of the sphere, and we link it to the three closest nodes. We then identify the nodes with number of neighbors between 1 and 2, identify the nodes with less than 3 neighbors that are the closest to them, and join them. We iterate and terminate the algorithm when no  more links were possible (at most one node has degree 2). Links longer than $70\%$ of the network geometrical diameter were removed, and the network was given a ``haircut", removing all bridges and possible isolated components. Finally, any nodes $k$ connected only to two other nodes $i$ and $j$ were removed and the links $\langle ik \rangle$ and $\langle  jk \rangle$ were replaced by a link $\langle  ij \rangle $. Except for the network ``haircut", the last few steps were meant to simplify the graph without loss of generality. 
This algorithm is meant to emulate the topology of network that grows and bifurcates from a central point, much like a growing vascular network.
An example of a network produced with this algorithm is shown in Fig.~\ref{fig:torus2}(b). The weight of the links in Fig.~\ref{fig:torus2}(b) has been assigned randomly.

The genus 2 torus was intended as a test case of our algorithm in graphs that are easily embeddable and where the generator basis vectors are already known, whereas the random 3D topology was intended as a test-case for naturally occurring graphs where typically the embedding is not known. The size of the graphs generated by these procedures fell into two classes: the $500$ class (size ranging from $N=478$ to $538$ nodes) and the $800$ class  (size ranging from $N=800$ to $960$ nodes).

\begin{figure}[t]
\includegraphics[width=0.95\linewidth]{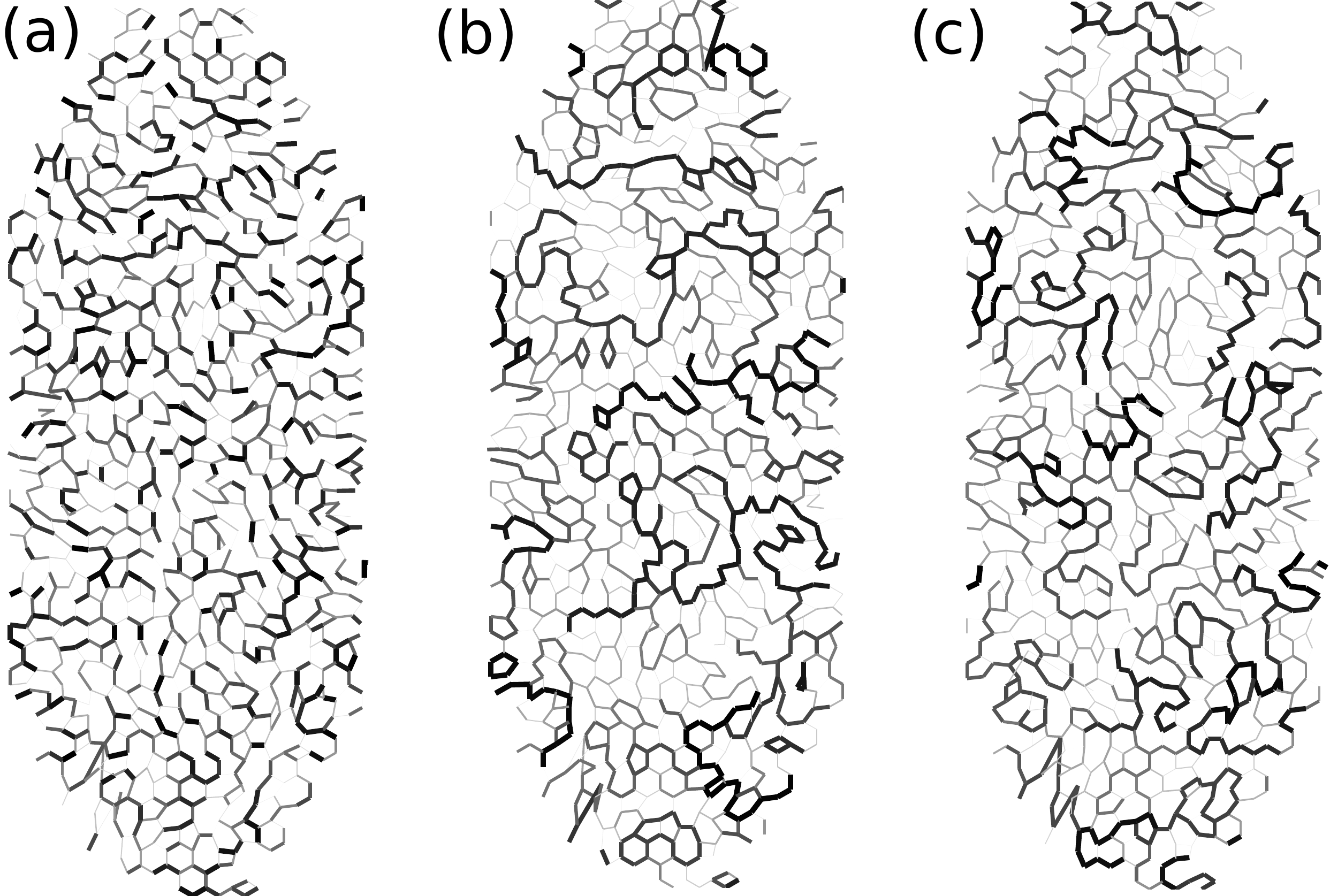}
\caption{\label{fig:weights} Examples of different edge weight assignment functions on a two-holed torus. (a) Random. (b) Lines, no self-avoidance. (c) Lines with SA. Note the high similarity between (b) and (c).}
\end{figure}

We used three main generating functions to assign the weights (in particular the weight order) to the underlying graph topologies. In the first function, termed {\it Random} in the following text, the weights are assigned at random (see e.g. Fig.~\ref{fig:weights}(a) ). This random, high disorder assignment, was intended as the baseline, control case to compare against our more ordered graphs.

For the second function, termed {\it Lines} (Fig.~\ref{fig:weights}(b)) we aimed to emulate a weight distribution with some long range order, a distribution that would generate linear segments randomly placed on the graph. The initial link $e_0$is chosen at random and assigned the highest weight. Starting from that link, we iteratively transverse a randomly chosen trail on the graph, assigning weights in descending order. When the trail reaches a node with outgoing links that have all been visited and have assigned weights, the trail terminates and a new trail is generated at another randomly selected edge with unassigned weight. At the initial stages of the weight assignment trails terminate mostly through self-intersections.

The third generating function was similar to the {\it Lines } model, with the addition of self avoidance (Fig.~\ref{fig:weights}(c)). Namely, the tip of the growing trail cannot intersect the last 40 nodes added to the trail. In addition, in this model, termed {\it SA Lines}, we implemented a length cutoff and a trail can be at most 12 links long.

Last, we explored weight assignments produced by a positive feedback adaptation algorithm described in Ref. \cite{Grawer2014} as a more biologically relevant test case. For more details about the adaptation algorithm the interested reader can consult \cite{Tero2010}. For completeness, we briefly describe the algorithm here and provide some further detail in the Appendix.

In our adaptive model, each pair of vertices of the network can act as a net current source and sink. The network carries the load from the source to the sink, and the conductivity of the links grows or shrinks according to the average flow through them (the average is being performed over all pair of vertices that act as sources and sinks). Starting from a random assignment of edge conductivities, the networks evolve and finally converge to a hierarchically organized architecture.

By appropriate non-dimensionalization we may reduce the control parameters for our dynamical, adaptive system to the load on the system, $\vartheta$, and $\gamma$, the sigmoidal exponent that controls the strength and sharpness of the feedback. In the simulations shown in this work, the transportation load is $\vartheta=10$, and $\gamma=0.3$ (model termed {\it Adapted 0.3}) and $\gamma=0.8$ (model termed {\it Adapted 0.8}). The underlying topology was random, and the simulation was initialized with various random conductivity value sets $C_{ij}(0)$.

\subsection{Quantifying the Results}

\begin{figure}[t]
\includegraphics[width=1.00\linewidth]{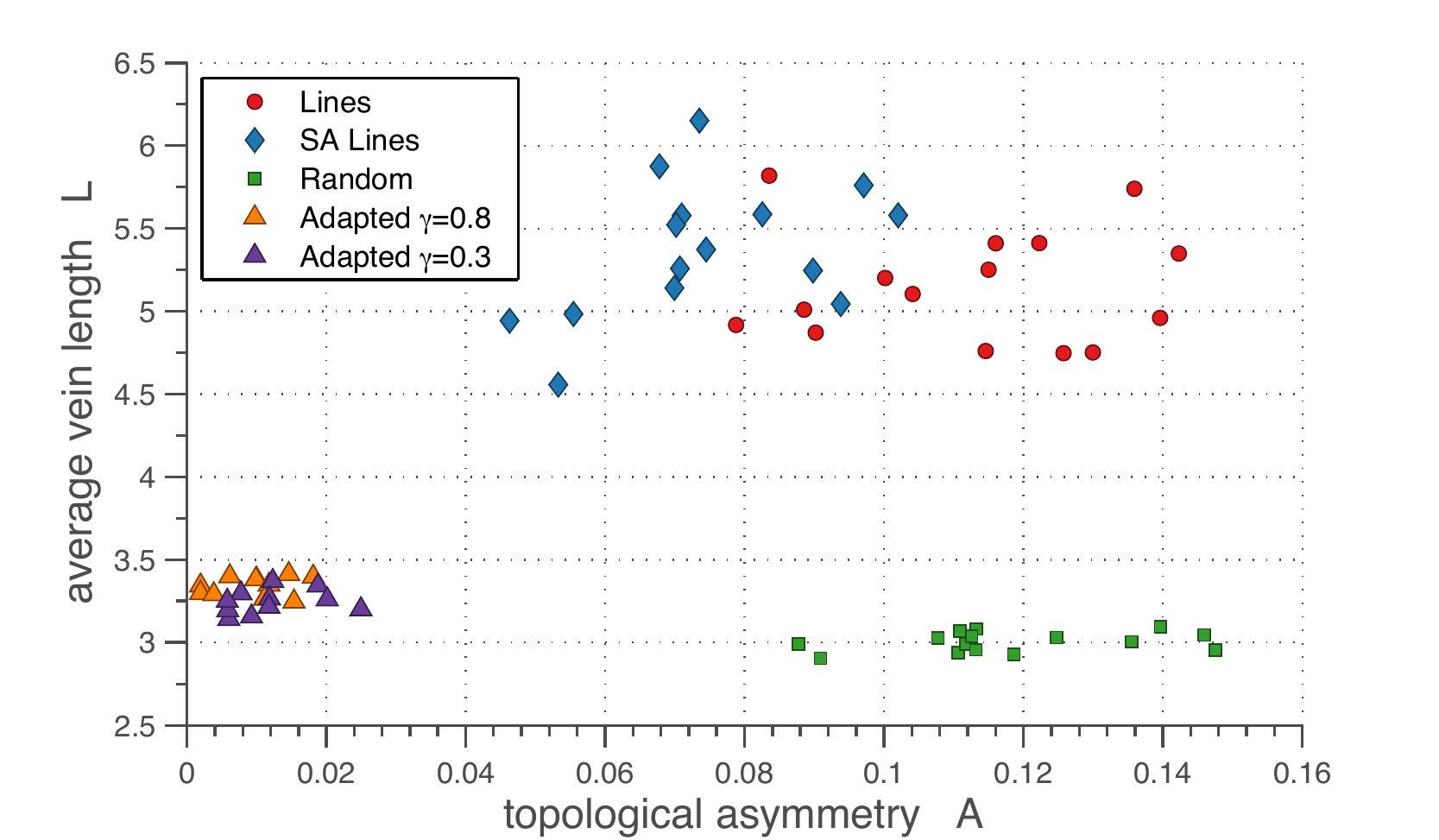}
\caption{\label{fig:sizes5003D} Average vein length versus topological asymmetry for various edge weight assignment models on random 3-regular graph topologies. The number of nodes on each graph was approximately $N \simeq 500$. Blue: SA Lines, red: Lines, green: random, orange: Adapted 0.8 and purple: Adapted 0.5.}
\end{figure}

We applied the cycle-coalescence algorithm to the example networks and calculated the topological asymmetry $A$ and the average vein length $L$, as described above. 
In this section we describe the results of the analysis and demonstrate the power of the algorithm.

In Fig.~\ref{fig:sizes5003D} for random 3D spatial network topologies we plot the topological asymmetry $A$ versus the average vein length $L$ for five different weight assignment models, ({\it Random}, {\it SA Lines}, {\it Lines}, {\it Adapted $\gamma=0.5$} and {\it Adapted $\gamma=0.8$}). Each dot  in Fig.~\ref{fig:sizes5003D} is a different realization of the weight distribution and underlying topology generation models. In all cases, the underlying graph topology was random and the network sizes were from the 500 class. We see that the topological asymmetry and underlying graph topology can distinguish all the models except the  {\it Adapted 0.8} and {\it Adapted 0.3}. Surprisingly, the algorithm can even distinguish the {\it Lines} from the {\it SA Lines}, two models that are only subtly different (see Fig.\ref{fig:weights}(a) and (b)). Note that $A$ alone would not be able to distinguish the {\it Lines} from the {\it Random} model. The frequent self-intersection of the ``veins" in the {\it Lines} model create an nested cycle architecture that is very similar to the {\it Random} model in a local level. However, examining $L$, a quantity that captures more long range information about the graph, we see that the two models are clearly distinguishable. Note also that $L$ alone would not be able to distinguish the {\it Lines} from the {\it SA Lines} model. In this case, the two models generate ``veins" that have approximately the same average length. However, locally the {\it SA Lines} model is more highly nested as it lacks the small cycles that are the results of self-intersections, and the models generate different values for $A$.

\begin{figure}[t]
\includegraphics[width=1.00\linewidth]{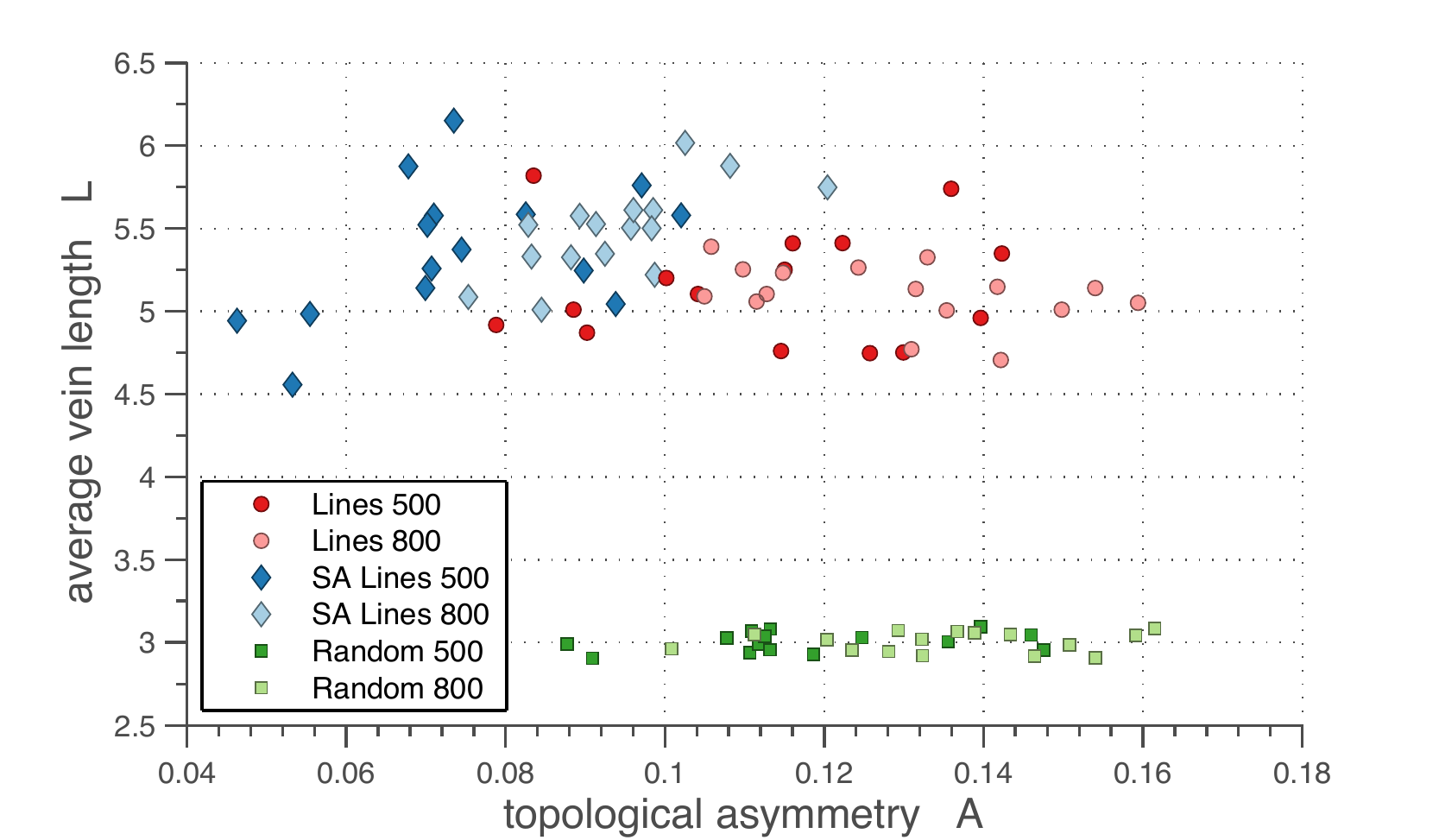}
\caption{\label{fig:sizes500and8003D} Average vein length $L$ versus topological asymmetry $A$ for various edge weight assignment models on random 3-regular graph topologies. The number of nodes on each graph was approximately $N\simeq 500$ or $N\simeq 800$. Blue: SA Lines 500, light blue: SA Lines 800, red: Lines 500, pink: Lines 800, green: Random 500, light green: Random 800. $L$ and $A$ are not sensitive to the graph size.}
\end{figure}

Although the topological asymmetry and the average vein length are generically size dependent quantities, their size dependence is not strong, at least for the models we examined.  In Fig.~\ref{fig:sizes500and8003D} we plot $L$ and $A$ for {\it Lines}, {\it SA Lines} and {\it Random}, for the two size classes, $500$ and $800$. The different size networks generated by the same weight assignment function are indistinguishable, despite the relatively large difference in graph size.

\begin{figure}[t]
\includegraphics[width=0.95\linewidth]{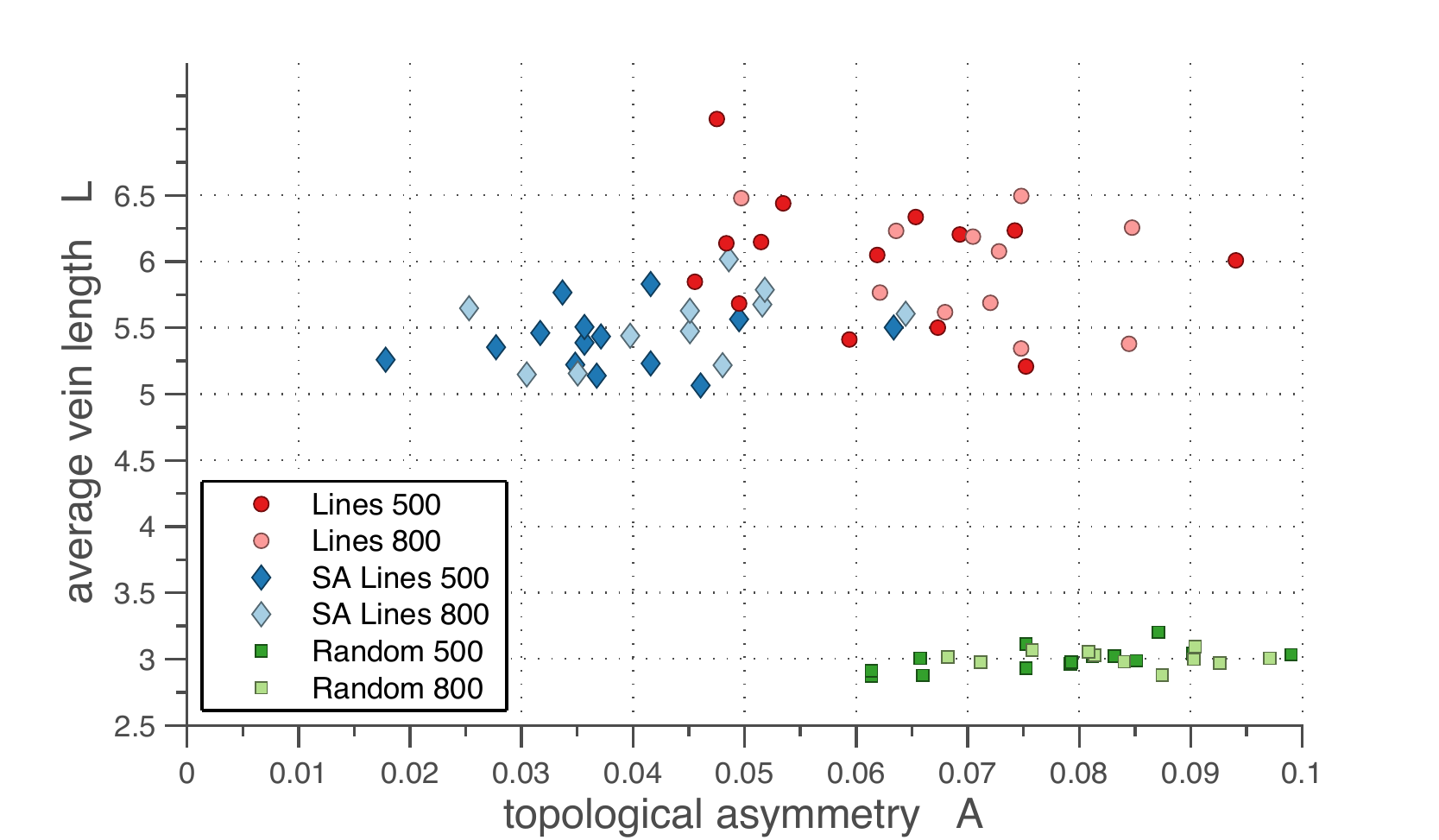}
\caption{\label{fig:sizes500and800Tori} Average vein length $L$ versus topological asymmetry $A$ for various edge weight assignment models on genus 2 toroidal topologies. The number of nodes on each graph was approximately $N\simeq 500$ or $N\simeq 800$. Blue: SA Lines 500, light blue: SA Lines 800, red: Lines 500, pink: Lines 800, green: Random 500, light green: Random 800. $L$ and $A$ are not sensitive to the graph size.}
\end{figure}

In all the above cases the underlying network topology was random. We repeated the same procedure for a two holed torus and see again that the results are qualitatively the same (Fig.~\ref{fig:sizes500and800Tori}). {\it Lines} and {\it SA Lines} are cleanly distinguishable but still obviously related, whereas {\it Random} is completely separate. Again, the 500 and 800 class networks remain identical.

\begin{figure}[t]
\includegraphics[width=0.95\linewidth]{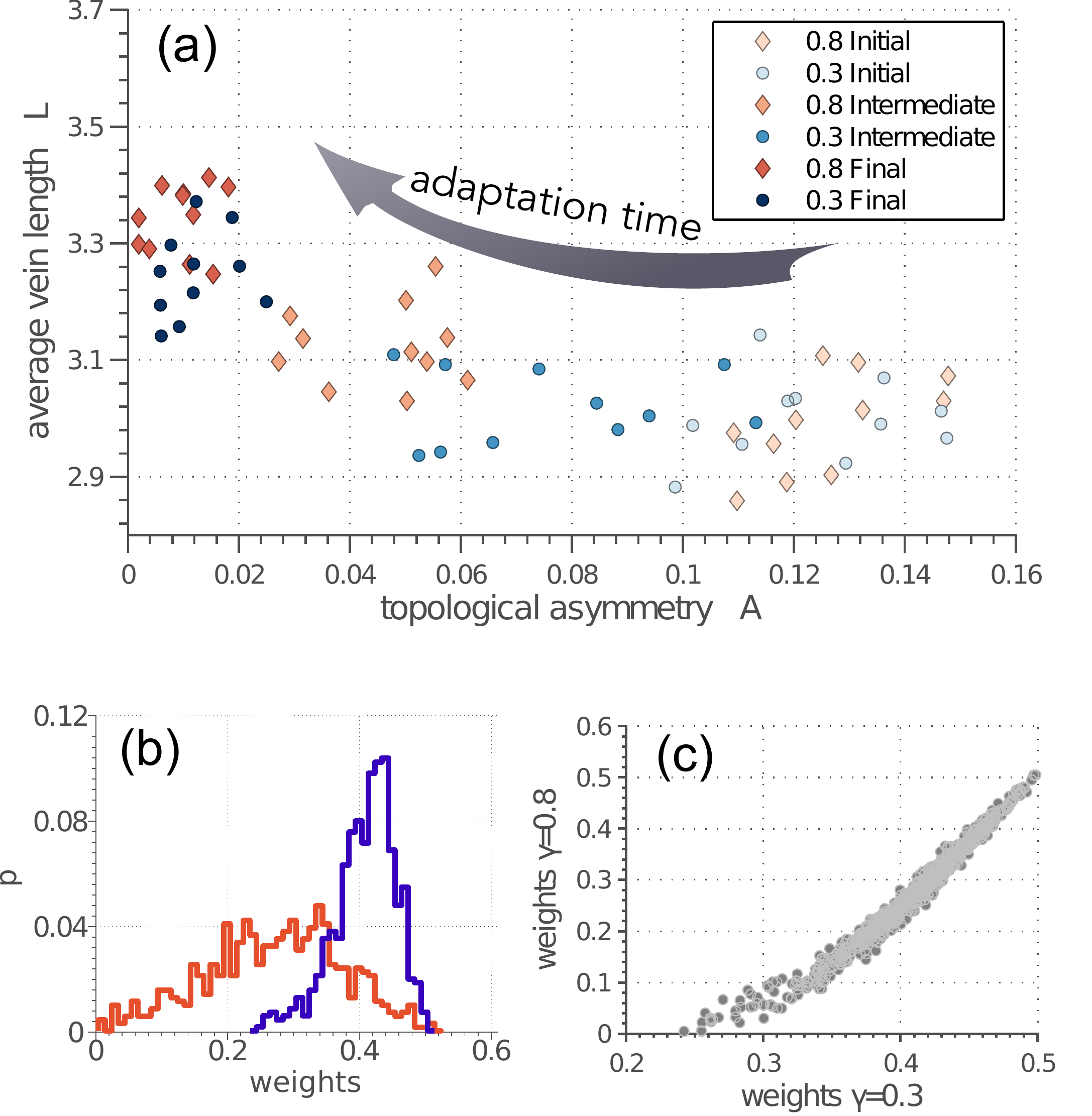}
\caption{\label{fig:Adaptation} (a) Average vein length $L$ versus topological asymmetry $A$ for initial, intermediate, and final states of an initially random-weighted network evolving under our adaptive model (b) Distribution of the edge weights in the final, adapted states for two different sigmoidal-feed-back control exponents, $\gamma$ (c) Structural correlations in the adapted weights from identical starting configurations for two different sigmoidal-feed-back control exponents, $\gamma$.}
\end{figure}

Finally, the cycle-coalescence algorithm can be used to investigate the evolving architecture of the adaptive model. Unsurprisingly, we discover a clear progression through our characteristic space from the unadapted, randomly weighted initial networks to final states with drastically lower topological asymmetry and slightly higher vein length (Fig. ~\ref{fig:Adaptation}(a)). Interestingly, however, despite the fact that varying the exponent in the sigmoidal of the adaptation's feedback function doesn't seem to affect where in our characteristic space the networks end up, by looking at intermediate stages of the adaptation one can clearly see that the higher exponent sample is much closer to its final state architecture and thus these networks are adapting ``faster" in a real sense. Despite the power evident in the choice of characterizing statistics we have chosen for the characteristic tree, we reiterate that not all information is being captured by these two probes. For example, the weight distribution in the final, adapted state is very different for the two different values of $\gamma$ (Fig. ~\ref{fig:Adaptation}(b)). Our method is only sensitive to the sort order of the weights and not the full weight distribution. The two fully adapted graphs $\gamma=0.3$ and $\gamma=0.8$ thus have very different weight distributions, but very similar architecture which could not have been guessed from information about the weights alone. Interestingly, starting from the same underlying topology and running the adaptation for the two different gammas, we see that $C^{\gamma=0.8}_{i,j}$ correlates strongly with $C^{\gamma=0.3}_{i,j}$ (Fig. ~\ref{fig:Adaptation}(c)), pointing to some kind of universal hierarchical organization for this system.

\section{Discussion}

We have presented a well-defined, statistically robust algorithm to characterize the structure and topology of weighted networks in three space dimensions. This method exactly matches known techniques in two dimensions and thus represents a natural extension and generalization of these earlier methods. Our quantification method can be used to understand the underlying architecture of 3D graphs in a way not possible before, as we have shown in the example of the adapting networks, where the $\gamma=0.3$ and $\gamma=0.8$ graphs may have wildly different weight distributions but are revealed to be structurally similar nonetheless. Indeed, working with traditional classification methods that attempt to leverage primarily edge-weight families has recently shown to be necessarily incomplete \cite{Barrat2004, DaFCosta2007}. Additionally, the inclusion of some kind of topological information to supplement that weight information is known to be desirable. Our cycle-coalescence algorithm represents just such a tool, but includes sensitivity to the geometry of the weight placement as well -- given networks with the same underlying topology and weights drawn from the same distribution it is easily possible to construct examples that functionally have very different architecture that no previously extant characteristic could distinguish.

Furthermore, our characteristic prescription is insulated against noise, as we have demonstrated only weak to non-existent sensitivity to simple variance in graph size, while the doubly statistical nature of the tool protects against tile misidentification. We point out that such misidentification must always remain a possibility for any algorithm that is expected to run in reasonable time for even moderately sized networks as the complete removal of the possibility of tile misidentification would represent a solution to the NP-hard graph genus problem.

It is our belief that this cycle-coalescence characterization of three dimensional networks will find wide applicability across many physical and biological representatives, hopefully uncovering new ways of thinking about these systems. From organ vasculature to neural networks, ant farms or hyphal networks to root networks of clonal colonies like quaking aspen, force networks in sand piles to airline routing, new descriptive and predictive modeling is possible.

\begin{acknowledgments}
Supported in part by NSF under grant number PHY-1058899. EK wishes to thank the Burroughs Wellcome Fund.
\end{acknowledgments}

\appendix

\section{Network Evolution Model}

The network evolution model we have chosen to use here is governed by a system of equations, describing transport, conservation and adaptation. The flow in the network is considered Hagen-Poiseuilleian laminar tube flow, so the transport equation for each edge is simply:
\begin{align}
    \label{eq:transportation}
    Q_{ij}^{kl} = C_{ij} \cdot \left(p_i^{kl}-p_j^{kl}\right).
\end{align}
$Q_{ij}^{kl}$ is the flow through the edge $\{i,j\}$, when node $k$ is a source and node $l$ a sink. The hydrostatic pressure difference $\Delta p_{ij}^{kl} := p_i^{kl}-p_j^{kl}$ along the tube between the pressures $p_i^{kl}$ and $p_j^{kl}$ defined at the nodes $i$ and $j$ acts as a potential difference from which the flow arises. The proportionality factor is the fraction of the tube's conductivity, denoted as $C_{ij}$.

Meanwhile, from flow conservation at each node, $j$, we have:
\begin{equation}
    \sum_{j,  \forall \ \{i,j\} \in \mathbb{E}}  Q_{ij}^{kl} = (\delta_{ik}-\delta_{il}) \cdot \zeta.
\end{equation} 
where $\mathbb{E}$ is the set of all edges.
For each node the sum of incoming flows must equal the sum of outgoing flows, unless the node is a source or sink which contributes an additional flow, $\zeta \ge 0$.

The ensemble averaged mean flow is:
\begin{align}
    \label{eq:meanflow}
    \langle|Q_{ij}|\rangle := \frac{1}{\frac{N\cdot(N-1)}{2}} \sum_{(k,l) \in \mathbb{P}} \left|Q_{ij}^{kl}\right|,
\end{align} 
where $\mathbb{P}$ is the set of all node pairs, and flows are considered equally in both directions.

Finally, we model the adaptation process with a differential equation describing the time evolution of the conductivities $C_{ij}=C_{ij}(t)$:
\begin{align}
    \label{eq:adaptation}    
    \frac{\dd C_{ij}(t)}{\dd t} = \beta \cdot f\left(\frac{\langle|Q_{ij}(t)|\rangle}{\epsilon}\right) - \alpha \cdot C_{ij}(t).
\end{align}
This equation features a positive, non-linear feedback term $\beta \cdot  f(\langle|Q_{ij}(t)|\rangle/\epsilon)$, that grows an edge's conductivity as a function of the scaled mean flow $\langle|Q_{ij}(t)|\rangle/\epsilon$ through itself. Balancing this term is a negative, exponential decay term $-\alpha \cdot C_{ij}(t)$. The parameters $\beta \ge 0$ and $\epsilon > 0$ scale the feedback and the flow through one edge; $\alpha \ge 0$ is the exponential decay parameter. The feedback function is sigmoidal, $f(x) = \frac{x^{\gamma}}{1+x^{\gamma}}$.

\bibliography{3DNetworkChar}

\end{document}